\documentclass[12pt]{article}
\usepackage{graphicx}
\usepackage{latexsym}
\usepackage{amscd}
\usepackage[utf8]{inputenc}
\usepackage[T1, T2A]{fontenc}
\usepackage[english]{babel}
\usepackage{amsmath}
\usepackage{amssymb}
\usepackage{indentfirst}
\usepackage[small,centerlast]{caption2}
\usepackage{longtable}

\usepackage{hyperref}
\usepackage{color}
\usepackage{xcolor}

\let\vec=\mathbf

\makeatletter
\renewcommand{\section}{\@startsection {section}{1}{\z@}%
	{-3.5ex \@plus -1ex \@minus -.2ex}%
	{2.3ex \@plus.2ex}%
	{\normalfont\Large}}
\renewcommand{\subsection}{\@startsection{subsection}{2}{\z@}%
	{-3.25ex\@plus -1ex \@minus -.2ex}%
	{1.5ex \@plus .2ex}%
	{\normalfont\large\itshape}}
\renewcommand{\subsubsection}{\@startsection{subsubsection}{3}{1em}%
	{-3.25ex\@plus -1ex \@minus -.2ex}%
	{-1.5em \@plus .2em}%
	{\normalfont\normalsize\bfseries}}

\makeatother

\usepackage[%
backend=biber,
bibencoding=utf8,
sorting=none,
style=gost-numeric,
autolang=other,
sortcites=true,
movenames = false,
maxnames = 3
]{biblatex}[2016/09/17]%
\DeclareFieldFormat[article]{title}{}
\DeclareFieldFormat[article]{titleaddon}{}

\DeclareFieldFormat[article]{journaltitle}{\mkbibemph{#1}}

\DefineBibliographyExtras{russian}{%

}
\DefineBibliographyExtras{english}{%

}   
\usepackage{xcolor}

\begin{document}

\centerline{\textbf{\Large ...}}

\begin{center}
	
	{\Large \textbf{Plasma Mechanism of Radio Emission Generation at the Bow Shock of the Exoplanet HD~189733b}}
	
	\medskip
	
	{\large \textbf{Kuznetsov A.A., Zaitsev V.V.}}
	
	\medskip
	\textit{A.V. Gaponov-Grekhov Institute of Applied Physics of the Russian Academy of Sciences, Nizhny Novgorod, Russia}\\

	\textit{kuznetsov.alexey@ipfran.ru}
	
\end{center}

\renewcommand{\abstractname}{}
\begin{abstract}
\noindent\small 
 
This study evaluates the possibility of efficient radio emission generation in the bow shock region of hot Jupiter–type exoplanets. As a source of energetic electrons, the shock drift acceleration mechanism at a quasi-perpendicular shock is proposed. Electrons reflected and accelerated by the shock propagate through the relatively dense stellar wind plasma and excite plasma waves; therefore, a plasma emission mechanism is considered as the source of the resulting radio waves. Using the bow shock of the hot Jupiter HD~189733b as a case study, the properties of the energetic electron beam, the excited plasma waves, and the resulting radio frequencies are estimated. An energy-based analysis is carried out to identify the range of stellar wind parameters for which radio emission from the bow shock of the exoplanet HD~189733b could be detectable by modern astronomical instruments.

\end{abstract}

\section{Introduction}

An important goal of modern radio astronomy is the detection of radio emissions from exoplanets, which could provide valuable insights into their plasma environments and the physical processes governing the evolution of stellar systems. The exoplanet HD~189733b is a typical and one of the most extensively studied representatives of the “hot Jupiter” class—giant planets comparable in size to Jupiter, located at close orbital distances (within 0.1 AU) from their host stars. This study explores the potential for radio emission generation in the bow shock region ahead of HD~189733b that could be sufficiently intense to be detected with current radio astronomical instruments. To this end, the kinetic energy density $W$ of the electron flux accelerated via a single reflection from the exoplanetary bow shock through the shock drift acceleration mechanism is estimated~\cite{Wu1984,Mann2005}. The stellar wind parameters considered here are such that the propagation of high-energy electrons leads to the development of the Langmuir instability, i.e., the excitation of plasma waves~\cite{Mikhailovsky1971,Treumann1997}. Consequently, the energy density $\mathcal{W}$ of these waves required for the generation of detectable radio emission via the plasma emission mechanism~\cite{Zaitsev1983} is also evaluated. By imposing the condition $W\gg\mathcal{W}$, the study estimates the region in the stellar wind parameter space where the detection of radio emission from the exoplanetary bow shock becomes energetically feasible. Additionally, the expected frequency range of the generated radio waves is assessed.

According to existing hydrodynamic and magnetohydrodynamic simulations of the HD~189733 system, the stellar wind velocity $v_s$ relative to the exoplanet may either exceed or fall below the fast magnetosonic speed $v_{ms}$ along the orbit of HD~189733b~\cite{Fares2017,Strugarek2022,Kavanagh2019,Odert2020,Rumenskikh2022}. When $v_{ms}>v_s$, the bow shock may not form~\cite{Zhilkin2019}. In this case, if the exoplanetary magnetic field is weak, the incoming stellar wind penetrates the plasmasphere, where a sufficient number of neutral particles exist. The differing collision frequencies of electrons and ions with neutrals lead to charge separation and, consequently, the emergence of an electric field component that accelerates electrons~\cite{Zaitsev2024}. In the present work, we consider the case $v_{ms}<v_s$, in which a bow shock forms ahead of the exoplanet. Electron acceleration is assumed to occur in the quasi-perpendicular region of the shock—where the angle $\theta$ between the shock normal and the interplanetary magnetic field approaches $90^\circ$. A fast magnetosonic shock is accompanied by magnetic field compression, effectively acting as a moving magnetic mirror. Electrons are reflected once from this mirror and accelerated in the process. This mechanism has been widely studied in the context of electron acceleration at Earth's bow shock~\cite{Holman1983,Wu1984,Liu2022}, interplanetary shocks~\cite{Yang2024}, and coronal shocks that give rise to type II radio bursts~\cite{Ball2001,Mann2005,Mann2018}.

By analogy with planets in the solar system, the electron cyclotron maser is often proposed as a mechanism for radio emission generated by accelerated electrons. However, this mechanism is effective only when the electron gyrofrequency significantly exceeds the plasma frequency, i.e., $\Omega_c\gg\omega_p$~\cite{Wu1979,Melrose1984,Louis2019}. For the stellar wind along the orbit of the exoplanet HD~189733b, the opposite inequality holds, $\Omega_c\ll\omega_p$, making the plasma emission mechanism potentially more effective~\cite{Zaitsev1983,Zaitsev2022}. This mechanism involves the generation of plasma waves by energetic electrons, followed by their conversion into electromagnetic radiation at the plasma frequency via scattering on plasma particles (Rayleigh scattering), or at twice the plasma frequency as a result of wave–wave interactions (Raman scattering). The possible frequency range of the resulting radio emission in this model is determined not by the magnetic field strength in the source region, but by the plasma density $n$ of the stellar wind at the exoplanetary bow shock.

In the case of Rayleigh scattering, which is most efficient when occurring on the ions of the background plasma, a maser effect may arise under certain conditions, manifesting as an exponential increase in the intensity of electromagnetic radiation with increasing plasma wave energy. In the case of Raman scattering, the maser effect is absent due to the decay of the electromagnetic wave into two plasma waves at high radio emission intensities. Nevertheless, the brightness temperature values in the source required to produce the observed radio flux can be comparable for both Rayleigh and Raman scattering~\cite{Zaitsev1983,Zheleznyakov1996,Zaitsev2023}.

Section~\ref{chapter2} provides a brief overview of the shock drift acceleration mechanism for electrons at a quasi-perpendicular shock. In Section~\ref{chapter3}, the kinetic energy density of fast electrons accelerated at the bow shock of HD~189733b is estimated for various stellar wind parameters. Section~\ref{chapter4} presents estimates of the plasma wave spectrum generated by energetic particles. Section~\ref{chapter5} briefly discusses the Rayleigh and Raman scattering mechanisms for the conversion of plasma waves into electromagnetic radiation and provides estimates for the plasma wave energy density required to produce a detectable radio emission flux at Earth. A comparison of the plasma wave energy density and the kinetic energy of accelerated electrons allows us to determine the region of stellar wind parameters within which the detection of radio emission resulting from electron acceleration at the bow shock of the exoplanet HD~189733b is energetically feasible.

\section{Shock Drift Acceleration of Electrons at a Quasi-Perpendicular Shock}
\label{chapter2}

\begin{figure}[h]
\includegraphics[width=8cm, height=6cm]{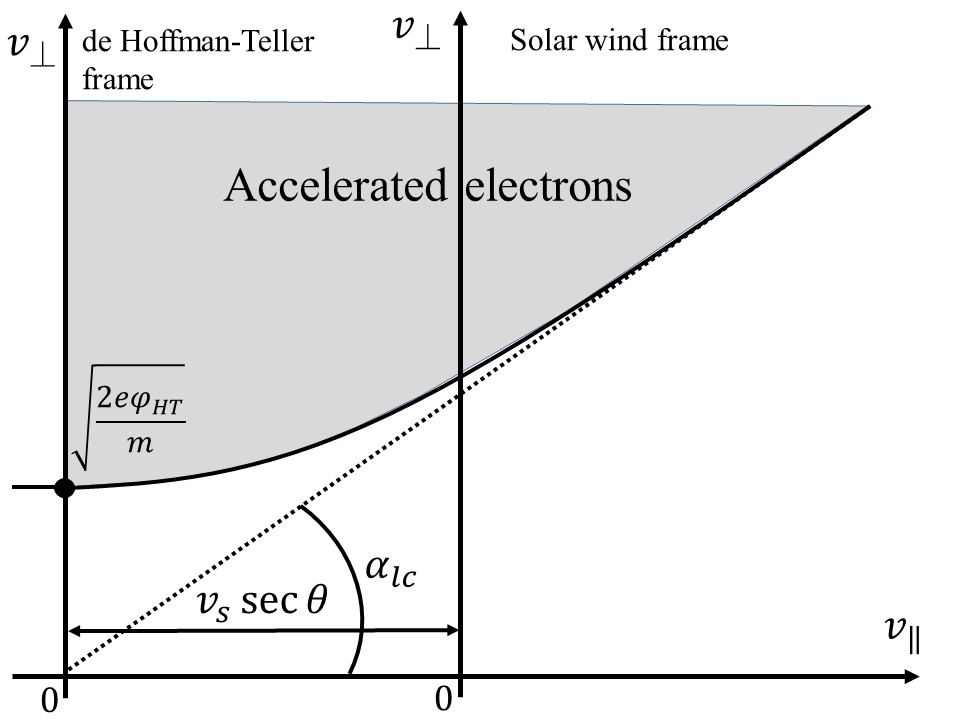}
\centering
\captionstyle{normal}
\caption{Schematic illustration of the stellar wind rest frame and the de Hoffmann–Teller frame. The shaded region indicates electrons that are reflected and accelerated.}
\label{ris:dHT}
\end{figure}

The consideration of the acceleration mechanism can be performed within the non-relativistic approximation, since the velocity of the reflected electron beam will later be shown to lie within the range of $1\cdot10^4\div3\cdot10^4~\textrm{km/s}$. The shock drift acceleration mechanism for electrons is conveniently described in the de Hoffmann–Teller reference frame (Fig.\ref{ris:dHT}), where the induced electric field vanishes due to the stellar wind velocity being aligned with the magnetic field in this frame~\cite{Ball2001,DeHoffmann1950}. Since a fast magnetosonic shock is accompanied by magnetic field compression with a characteristic scale of inhomogeneity much larger than the electron gyroradius, the shock acts as a magnetic mirror moving along the field lines with a velocity of $v_s\sec\theta$. In this case, electrons that do not enter the loss cone can be reflected once by the shock and accelerated. The components of the electron velocity parallel to the magnetic field before and after reflection, $v_{i,\|}$ and $v_{r,\|}$ respectively, are related by the expression (\ref{eq:reflection}).

\begin{equation}
    v_{r,\|}=2v_s\sec\theta-v_{i,\|},
    \label{eq:reflection}
\end{equation}
where $v_s=\sqrt{v_{sw}^2+v_{orb}^2}$ is the velocity of the stellar wind relative to the shock, equal to the root mean square of the stellar wind velocity at the exoplanet's orbit $v_{sw}$ and the orbital velocity of HD~189733b $v_{orb}$, which is no longer negligible as it is for the Solar System planets~\cite{Vidotto2010}. From equation (\ref{eq:reflection}), it is evident that the velocity gain during the reflection of an individual electron is proportional to the stellar wind velocity and increases as the angle approaches $90^\circ$. The component of velocity perpendicular to the magnetic field, $v_{\perp}$, remains unchanged upon reflection.

For solar wind electrons, it is known that their velocity distribution at high energies follows a power-law decay~\cite{Echim2010,Dudik2017,Kuznetsov2022}, and their temperature may differ from that of the ions, both higher and lower~\cite{Shi2023}. However, due to a lack of detailed data and for simplicity, it is assumed that the electrons in the initial plasma of the stellar wind of HD~189733 at the exoplanetary orbit have a Maxwellian isotropic velocity distribution (\ref{eq:maxw}) and the same temperature $T$ as the ions:

\begin{equation}
    f_{bkg}\left(v_\perp,v_\|\right)=\frac{1}{(2\pi v_{th})^{3/2}}\exp\left(-\frac{v_\perp^2+v_\|^2}{2v_{th}^2}\right),
    \label{eq:maxw}
\end{equation}

where $v_{th}=\sqrt{k_bT/m_e}$ is the thermal velocity of the stellar wind electrons. In this case, the distribution of reflected electrons takes the form:
\begin{equation}
    f_{acc}\left(v_\perp,v_\|\right)= \frac{\Theta(v_\|-v_s\sec\theta)\Theta\left(v_\perp-v_{\perp,lc}\right)}{(2\pi v_{th})^{3/2}}\exp\left(-\frac{v_\perp^2+(v_\|-2v_s\sec\theta)^2}{2v_{th}^2}\right),
    \label{eq:f_accel}
\end{equation}

where $\Theta(x)$ is the Heaviside step function such that $\Theta(x)=1$ for $x>0$ and $\Theta(x)=0$ for $x<0$, and $v_{\perp,lc}=\tan\alpha_{lc}\sqrt{(v_\|-v_s\sec\theta)^2+V_e^2}$, where the effective velocity $V_e=\sqrt{\frac{2e\phi_{HT}}{m_e}}$ is introduced, associated with the cross-shock potential $\phi_{HT}$. The pitch angle $\alpha_{lc}$ is defined as $\alpha_{lc}=\arcsin\left[\left(B_1/B_2\right)^{1/2}\right]$, where $B_1$ and $B_2$ are the magnetic field strengths before and after the shock, respectively. In the MHD approximation, based on the Rankine–Hugoniot relations~\cite{Priest1982}, the mirror ratio $X=B_1/B_2$ for a quasi-perpendicular shock satisfies the equation~(\ref{eq:X}):

\begin{equation}
   aX^{-2}+b X^{-1}+c=0,
    \label{eq:X}
\end{equation}
where
\begin{equation}
a=2-\gamma,~b=\left(\frac{2c_s^2}{\gamma v_a^2}+(\gamma-1)\frac{v_s^2}{\gamma v_a^2}+1\right)\gamma ,~c=-(\gamma+1)\frac{v_s^2}{ v_a^2},
\end{equation}

with the Alfvén velocity defined as $v_a^2=B^2/{4\pi nm_i}$ and the sound speed as $c_s^2=\gamma k_bT/m_i$, where the adiabatic index is $\gamma=5/3$. In a dense stellar wind, satisfying $b^2\gg4ac$ and $v_a^2\ll v_s^2$, the mirror ratio becomes independent of the wind plasma density and magnetic field strength~(\ref{eq:approx_ratio}):
\begin{equation}
    X\approx-\frac{b}{c}\approx\frac{2c_s^2+(\gamma-1)v_s^2}{(\gamma+1)v_s^2}.
    \label{eq:approx_ratio}
\end{equation}
Due to the difference in inertia between protons and electrons across the shock, a cross-shock potential $\phi_{HT}$ is established~\cite{Goodrich1984}, given by:
\begin{equation}
    e\phi_{HT}=\frac{\gamma}{\gamma-1}k_bT.
    \label{eq:HT_potential}
\end{equation}
The velocity distribution of the accelerated particles~(\ref{eq:f_accel}) belongs to the family of shifted loss-cone distributions~\cite{Wu1984}. It is nonzero only when two conditions are satisfied: the particle velocity exceeds the shock velocity along the magnetic field, $v_\|>v_s\sec\theta$, and the particle does not fall into the loss cone, $v_\perp>v_{\perp,lc}$.

\section{Electron Acceleration at the Bow Shock of the Exoplanet HD~189733b}
\label{chapter3}

The star HD~189733 belongs to the class of orange dwarfs, with a radius of $R_{star} \approx 0.76R_\odot$, and hosts a hot Jupiter-type exoplanet on an orbit of approximately $\approx 9R_s$. At such distances, the stellar wind becomes significantly azimuthally inhomogeneous in the plane of the planetary orbit due to the non-uniform outflow of plasma from the stellar surface. As shown by MHD simulations~\cite{Fares2017,Strugarek2022,Odert2020}, at least part of the planet’s orbit lies within a region where the relative velocity between the stellar wind and the exoplanet, $v_s$, exceeds the fast magnetosonic speed $v_{ms}$: 
\begin{equation}
    v_s=\sqrt{v_{sw}^2+v_{orb}^2}>v_{ms}=\sqrt{v_a^2+c_s^2}.
    \label{eq:condition}
\end{equation}
Inequality~(\ref{eq:condition}) is the condition for the formation of a fast magnetosonic shock~\cite{Priest1982}, which is necessary for the shock drift acceleration of electrons (see Section~\ref{chapter2}).

The shape, position, and nature of the exoplanetary bow shock remain the subject of active investigation and depend strongly on the parameters of the stellar wind~\cite{Vidotto2010,Llama2013,Bourrier2013,Zhilkin2019,Rumenskikh2022}. Depending on the wind intensity, the shock may form at distances ranging from 3 to 20 planetary radii from the center of HD~189733b. The minimum distance occurs during a coronal mass ejection from the host star, corresponding to the stellar wind parameter set N4~\cite{Odert2020}. Parameter set N3 comprises stellar wind conditions most favorable for shock drift acceleration among those encountered along the exoplanet's orbit in MHD simulations~\cite{Odert2020}. Set N2 consists of averaged stellar wind parameters based on the data from~\cite{Kavanagh2019}. Finally, set N1 uses parameters from a hydrodynamic simulation~\cite{Rumenskikh2022}, augmented by a relatively weak magnetic field $B = 0.01$~G, sufficient for the formation of a magnetosonic shock and a moderately wide loss cone, allowing for efficient acceleration.

In this section, we first estimate the energy density of accelerated electrons, normalized to their thermal energy (\ref{eq:density_electron_energy}), for the selected parameter sets, and then analyze its dependence on stellar wind number density $n$, velocity $v_{sw}$, and magnetic field strength $B$ at a characteristic temperature of $T = 1.5\cdot10^6$~K.

\begin{table}[h!]
\centering
\begin{tabular}{ |p{3cm}||p{2cm}|p{2cm}|p{2cm}|p{2cm}| }
 \hline
  Set & N1~\cite{Rumenskikh2022} & N2~\cite{Kavanagh2019} & N3~\cite{Odert2020}& N4~\cite{Odert2020} \\
 \hline
  $v_{sw}$,km/s   & $240$ & $235$  & $480$& $1000$ \\
 \hline
  $n,\textrm{cm}^{-3}$   & $10^5$ & $4\cdot10^6$  & $4.8\cdot10^5$& $4.4\cdot10^6$\\
 \hline
  $T,K$  & $10^6$ & $10^6$ & $2.2\cdot10^6$& $2\cdot10^6$ \\
 \hline
  $B$, G  & $0.01$ & $0.062$ & $0.022$& $0.1$ \\
 \hline
   $L$, $R_p$  & $5$ & $4.5$ & $4.5$& $3$ \\
 \hline
\end{tabular}
\caption{Stellar wind parameter sets used in the estimates}
\label{tab:param}
\end{table}

\begin{figure}[h]
\includegraphics[width=8cm, height=6cm]{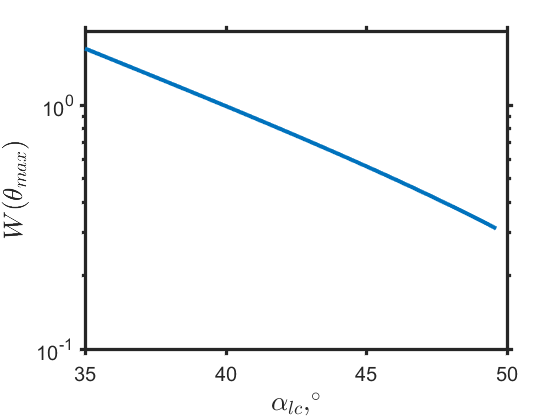}
\centering
\captionstyle{normal}
\caption{Dependence of the dimensionless maximum energy density of accelerated electrons $W(\theta_{max})$ on the pitch angle $\alpha_{lc}$.}
\label{ris:alpha_lc}
\end{figure}

By integrating the velocity distribution of accelerated particles~(\ref{eq:f_accel}) over velocity space, one can obtain expressions for their number density $n_{acc}$ and the characteristic longitudinal $v_{b,\|}$ and transverse $v_{b,\perp}$ velocities of the beam of reflected electrons, depending on the angle $\theta$ between the magnetic field and the shock normal~\cite{Mann2005}.
\begin{equation}
    \frac{n_{acc}(\theta)}{n}=\exp\left(-\frac{v_s^2\sec^2\theta\sin^2\alpha_{lc}+V_e^2\tan^2\alpha_{lc}}{2 v_{th}^2}\right)\cdot\frac{\cos\alpha_{lc}}{2}\left(1+\mathrm{erf}\left(\frac{\sqrt{2}v_s\sec\theta\cos\alpha_{lc}}{v_{th}}\right)\right),
    \label{eq:dens_acc}
\end{equation}
\begin{equation}
    v_{b,\|}(\theta)=v_{s}\sec\theta\left(1+\cos^2\alpha_{lc}\right),
\end{equation}
\begin{equation}
    v_{b,\perp}(\theta)=\tan\alpha_{lc}\sqrt{(v_{b,\|}-v_s\sec\theta)^2+V_e^2},
\end{equation}
The dimensionless energy density of the accelerated electrons is defined as
\begin{equation}
W(\theta)\approx{m_en_{acc}}\left(v_{b,\|}^{2}+v_{b,\perp}^{2}\right)/2nk_bT
\label{eq:density_electron_energy}
\end{equation}
and for $v_s \sin\alpha_{lc} \ll \sqrt{2}v_{th}$ it reaches a maximum at an angle $\theta_{max}$ close to $90^\circ$~(\ref{eq:max_angle}). Thus, the angular range over which the energy density remains comparable to the maximum is estimated as $\Delta\theta \approx 90^\circ - \theta_{max}$, and for the chosen parameters it is approximately $5^\circ \div 15^\circ$. The maximum value of the energy density $W(\theta_{max})$ and the corresponding angle $\theta_{max}$ are obtained under the assumption $\frac{\sqrt{2}v_s \sec\theta \cos\alpha_{lc}}{v_{th}} \gg 1$~(\ref{eq:max_en}--\ref{eq:max_angle}). The value of $W(\theta_{max})$ depends solely on the pitch angle $\alpha_{lc}$~(\ref{eq:max_en}), and through its definition, on the stellar wind parameters. It decreases monotonically with increasing $\alpha_{lc}$ (see Fig.\ref{ris:alpha_lc}). For the selected parameter sets, the pitch angle increases from $32^\circ$ to $43^\circ$, resulting in a one-order-of-magnitude decrease in the peak energy density $W(\theta_{max})$ (Table~\ref{tab:results}).

\begin{equation}
W(\theta_{max})\approx \cos(\alpha_{lc})\left(\left(\frac{1+\cos^2\alpha_{lc}}{\sin\alpha_{lc}}\right)^2+\tan^2\alpha_{lc}\left(\cos^2\alpha_{lc}/\tan^2\alpha_{lc}+\frac{\gamma}{\gamma-1}\right)\right)\exp\left(-1-\frac{\gamma\tan^2\alpha_{lc}}{2\gamma-2}\right)
\label{eq:max_en}
\end{equation}
\begin{equation}
\theta_{max}=\arccos\sqrt{\frac{v_s^2\sin^2\alpha_{lc}}{2v_{th}^2}}
\label{eq:max_angle}
\end{equation}

\begin{table}[h!]
\centering
\begin{tabular}{ |p{3cm}||p{2cm}|p{2cm}|p{2cm}|p{2cm}| }
 \hline
  Set & N1 & N2 & N3 & N4 \\
 \hline
  $\alpha_{lc},^\circ$  & $42$ & $43$  & $37$& $32$ \\
   \hline
  $v_s/v_{ms}$  & $1.95$ & $1.92$  & $2.6$ & $5.1$ \\
 \hline
  $n_{acc}(\theta_{max}),~\textrm{cm}^{-3}$ & $2\cdot10^3$ & $1.7\cdot10^5$  & $4\cdot10^4$ & $5.7\cdot10^5$ \\
 \hline
  $v_{b,\|}(\theta_{max})$, km/s & $1.5\cdot10^4$  & $10^4$  & $2.06\cdot10^4$ & $2.4\cdot10^4$ \\
   \hline
  $v_{b}(\theta_{max})$, km/s & $1.8\cdot 10^4$& $1.34\cdot10^4$  & $2.35\cdot10^4$& $2.6\cdot10^4$\\
 \hline
  $W(\theta_{max})$ & $0.26$ & $0.26$  & $0.68$ & $1.43$ \\
   \hline
\end{tabular}
\caption{Estimated efficiency of electron acceleration at the bow shock of the exoplanet HD~189733b.}
\label{tab:results}
\end{table}

\begin{figure}[h]
\includegraphics[width=8cm, height=6cm]{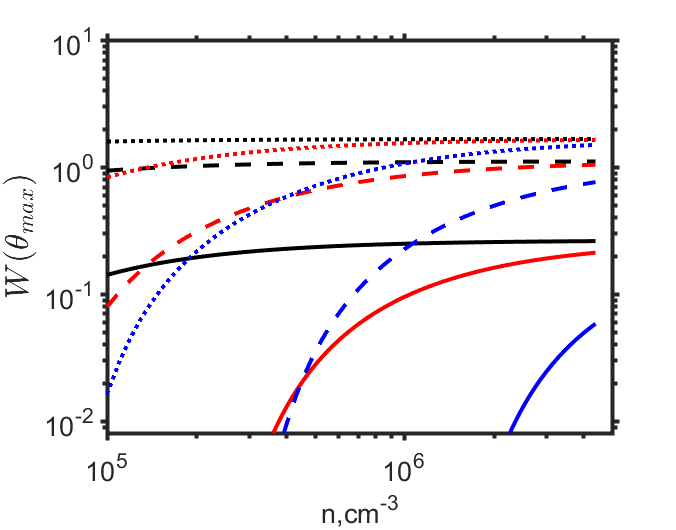}
\centering
\captionstyle{normal}
\caption{Dependence of the energy density of accelerated electrons $W(\theta_{max})$ on the number density $n$ at temperature $T = 1.5 \cdot 10^6$ K, and magnetic field strengths $B = 0.01$ G (black), $B = 0.04$ G (red), $B = 0.1$ G (blue), for stellar wind velocities of $v = 250$ km/s (solid), $v = 500$ km/s (dashed), and $v = 1000$ km/s (dotted).}
\label{ris:W_zav_B2_V2}
\end{figure}

In the limit of low Alfvén velocity, $v_a \ll v_s$, the pitch angle $\alpha_{lc}$ depends only on the temperature and bulk velocity of the stellar wind~(\ref{eq:approx_ratio}). Therefore, at high plasma density $n$, the energy density $W(\theta_{max})$ approaches an asymptotic value (Fig.~\ref{ris:W_zav_B2_V2}). As the magnetic field strength $B$ increases and the stellar wind velocity $v_s$ decreases, the maximum energy density $W(\theta_{max})$ diminishes.

\section{Generation of Plasma Waves}
\label{chapter4} 
The electrons reflected from the shock return to the unmagnetized stellar wind plasma, where $\omega_p \gg \omega_c$. Therefore, at the angle between the shock normal and the magnetic field $\theta = \theta_{max}$, which is close to the optimal value, the total velocity distribution function takes the form:

\begin{equation}
f\left(v_\perp,v_\|,\Theta_{max}\right)=f_m\left(v_\perp,v_\|\right)+f_{acc}\left(v_\perp,v_\|,\Theta_{max}\right),
\label{eq:full_fr}
\end{equation}
Although the total distribution function is axially symmetric, its two-dimensional cross-section in the $(v_\perp, v_\|)$ coordinates can be interpreted as a high-energy beam of accelerated electrons propagating through the warm and dense stellar wind plasma. Expressions for both components of the beam drift velocity $v_b$ are provided in Section~\ref{chapter3}. The thermal spread of particles in the beam is, strictly speaking, anisotropic. However, since their source is the same stellar wind, it will be assumed in the estimates below to be equal to the thermal speed of the background plasma $v_{th}$. Given that $v_b \gtrsim \sqrt{3}v_{th}$ for all considered parameter values, there exists a range of particle velocities for which the inequality $\partial f / \partial |v| > 0$ holds. As a result, beam-plasma instability~\cite{Mikhailovsky1971,Treumann1997} efficiently excites Langmuir (plasma) waves with plasma frequency $\omega_p$ given by:
\begin{equation}
    \omega_p^2\approx\omega_L^2+3k^2v_{th}^2,
    \label{eq:plasma_freq}
\end{equation}
where $\omega_L^2=4\pi ne^2/m_e$. The velocity interval in which the distribution function satisfies the condition $\partial F/\partial |v|>0$ can be roughly estimated as $v\in \left(v_b-v_{th};v_b\right)$. Using the resonance condition $\omega_p\approx kv$, this yields an estimate~(\ref{eq:est_k}) for the range of unstable wavenumbers $k$ (and corresponding plasma wave frequencies $f_p$) of the excited Langmuir waves: 

\begin{equation}
    k_p\in\left(k_{p,min};k_{p,max}\right)=\left(\frac{\omega_L^2}{(v_b-v_{th})^2-3v_{th}^2};\frac{\omega_L^2}{v_b^2-3v_{th}^2}\right).
    \label{eq:est_k}
\end{equation}
From this, one can estimate the characteristic wavenumber $\langle k_p\rangle$ of the spectrum and the width of the unstable wavenumber range $\Delta k_p$: 
\begin{equation}
    \langle k_p\rangle\approx\frac{k_{p,min}+k_{p,max}}{2},~~~~\Delta k_p\approx k_{p,max}-k_{p,min}
    \label{eq:est_width_charakt}
\end{equation}
For Langmuir instability to develop, its growth rate must exceed the effective electron-ion collision frequency $\nu_{ei}$, which is defined by ~\cite{Zheleznyakov1996} as: 
\begin{equation}
    \nu_{ei}=\sqrt{\frac{8\pi}{m_e}}\frac{e^4n}{\left(k_bT\right)^{3/2}}\ln\left(0.37\frac{k_bT}{e^2n^{1/3}}\right)
    \label{eq:freq_ei_coll}
\end{equation}

For the stellar wind parameters used in this work, the condition $\gamma_{max} \gg \nu_{ei}$ is satisfied. It should also be noted that due to scattering of Langmuir waves by the stellar wind particles, their spectrum becomes isotropized. Therefore, without loss of generality, we may assume for simplicity that the spectrum of Langmuir waves is isotropic, with spectral energy density $W_k$. In that case, the total energy density of plasma waves is given by $W_p = \int W_kdk$.

\section{Conversion of Plasma Waves into Radio Emission}
\label{chapter5}
Plasma waves excited at the plasma frequency $\omega_p$ are converted into electromagnetic radiation through scattering off stellar wind particles (Rayleigh scattering). The corresponding conservation law is given by:

\begin{equation}
    \omega_t-\omega_p=(\vec{k}_t-\vec{k}_p)\vec{v},
    \label{Eq:rayleigh_scat}
\end{equation}

where $\omega_t$ and $\vec{k}_t$ are the frequency and wave vector of the electromagnetic wave, and $\vec{v}$ is the velocity of the scattering particle. Rayleigh scattering leads to radio emission at a frequency $\omega_t \approx \omega_p$.

Nonlinear interaction of two plasma waves (Raman scattering) described by:
\begin{equation}
    \omega_p+\omega_p'=\omega_t,~\vec k_p+ \vec k_p'=\vec k_t 
\end{equation}
generates radio emission at the second harmonic of the plasma frequency, $2\omega_p$.

In radio astronomy, the intensity of radiation from cosmic sources is characterized by the brightness temperature $T_b$. This quantity is related to the radio flux $F$, measured at a distance $R_{se}$ from the source, by the following expression~\cite{Zheleznyakov1996}:
\begin{equation}
    F=\frac{2\omega_t^2k_bT_b}{(2\pi c)^2}\frac{R_s^2}{R_{se}^2},
    \label{eq:flux_common}
\end{equation}
where $R_s$ is the characteristic size of the source within the line of sight. Since the origin of the radio emission lies in the electrons accelerated at the quasi-perpendicular exoplanetary bow shock, the value of $R_s$ is taken to be the characteristic linear size of the shock. As shown by both estimates and numerical modeling, the location, shape, and type of the exoplanetary shock strongly depend on the stellar wind parameters~\cite{Zhilkin2019}. Therefore, its linear size is conservatively estimated as $R_s \approx L R_p \sin(\Delta\theta)$, where $R_p$ is the radius of the exoplanet HD~189733b.

The variation of brightness temperature $T_b$ along the propagation path is described by the radiative transfer equation:
\begin{equation}
    \frac{dT_b}{dl}=a_i-(\mu_{Ni}+\mu_{C})T_b
    \label{eq:transfer}
\end{equation}
where index $i=1$ corresponds to Rayleigh scattering, and $i=2$ to Raman scattering. In equation~(\ref{eq:transfer}), the coefficient $a_i$ represents spontaneous scattering; $\mu_{N1}$ is the coefficient of induced scattering of plasma waves into electromagnetic waves; $\mu_{N2}$ denotes the coefficient of nonlinear absorption of electromagnetic waves; and $\mu_C$ is the absorption coefficient due to Coulomb collisions.

In subsections~\ref{chapter5.1} and~\ref{chapter5.2}, using equations~(\ref{eq:flux_common}) and~(\ref{eq:transfer}), a relation will be derived between the dimensionless energy density of plasma waves $\mathcal{W}_i = W_p / (n k_b T)$ and the radio flux $F$ detectable on Earth. Then, in subsection~\ref{chapter5.3}, the obtained relations will be used to estimate the minimum plasma wave energy density $\mathcal{W}_i$ required for detection by modern radio telescopes, both for Rayleigh and Raman scattering scenarios.

Finally, the condition $W \gg \mathcal{W}_i$ will be used to evaluate the stellar wind parameters under which radio emission from the region of the exoplanetary bow shock can be energetically detectable.

\subsection{Rayleigh scattering}
\label{chapter5.1} 
If the coefficient of induced scattering $\mu_{N1}$ is negative and sufficiently large in absolute value $\left(|\mu_{N1}| > \mu_C\right)$, the total coefficient in the right-hand side of equation~(\ref{eq:transfer}) becomes negative. In this case, the induced effect in Rayleigh scattering becomes dominant, and the brightness temperature $T_b$ grows exponentially with increasing energy density of plasma waves. This phenomenon is known as the plasma maser. The efficiency of maser amplification of electromagnetic waves is determined by the optical depth of the source $\tau$:

\begin{equation}
    \tau=\int_0^{R_s}|\mu_{N1}+\mu_{C}|dl
\end{equation}

In the case of Rayleigh scattering, at appropriate plasma wave energy densities, the optical depth $\tau$ can reach high values, enabling substantial radio wave fluxes from the source~\cite{Zaitsev2022}.

As mentioned above, the conversion process~(\ref{Eq:rayleigh_scat}) is most efficient when plasma waves are scattered by ions of the background plasma. In such scattering events, the frequency shift per act is negligible. In this case, the scattering is differential: the spectral width of the plasma waves exceeds the width of the kernel of the integral equation describing the induced scattering~\cite{Zheleznyakov1996}. The coefficients of spontaneous ($a_1$) and induced ($\mu_{N1}$) scattering used in (\ref{eq:transfer}) for an isotropic plasma wave spectrum are given by~\cite{Tsytovich1977}:
\begin{equation}
    a_1=\frac{\pi}{36}\frac{\omega_L^3W_k}{v_gnv_{th}^2k_p}
    \label{eq:a_1_koef}
\end{equation}
\begin{equation}
\mu_{N1}=-\frac{\pi}{108}\frac{m_e\omega_L^3}{m_iv_gnk_bTv_{th}^2k_p}\frac{\partial }{\partial k_p} \left(k_pW_k\right)
    \label{eq:mu_N_1_koef}
\end{equation}
The collisional absorption coefficient ($\mu_{C}$) is defined by~\cite{Ginzburg1970}:
\begin{equation}
    \mu_{C}=\frac{\omega_L^2\nu_{ei}}{\omega_t^2v_g}
\label{eq:mu_C_1_coeff}
\end{equation}
In Eqs.(\ref{eq:a_1_koef})–(\ref{eq:mu_C_1_coeff}), $v_g = c \sqrt{1 - \omega_L^2 / \omega_t^2} = \sqrt{3} c k_p v_{th} / \omega_p$ is the group velocity of electromagnetic waves with frequency $\omega_t^2 \approx \omega_p^2 = \omega_L^2 + 3 k_p^2 v_{th}^2$ in the region of plasma wave generation, $\omega_t = 2\pi f_t$ is the frequency of the electromagnetic wave, and $\nu_{ei}$ is the electron-ion collision frequency defined by Eq.(\ref{eq:freq_ei_coll}).

When solving the transfer equation  (\ref{eq:transfer}), it is convenient to change the integration variable from spatial coordinate $l$ (along the radiation propagation path) to the plasma wave vector $k\_p$, integrating from $k_{p,min}$ to $k_{p,max}$. Since $\omega_p = const$, the relation between $dl$ and $dk_p$ is:  

\begin{equation}
    dl=6L_n\frac{v_{th}^2}{\omega_p^2}k_pdk_p,
\end{equation}
where $L_n = |n (dn/dl)^{-1}|$ is the characteristic scale of plasma inhomogeneity, estimated as $L_n \approx R_s \approx L R_p \sin(\Delta \theta)$. Substituting into equation~(\ref{eq:mu_N_1_koef}), the optical depth of induced scattering of plasma waves into electromagnetic radiation, $\tau_{N1}$, is given by:
\begin{equation}
    \tau_{N1}=\int_0^{R_s}\mu_{N1}dl\approx-\frac{\pi}{18\sqrt{3}}\frac{m_e\omega_L\langle v_{ph} \rangle}{m_i cv_{th}}R_s\mathcal{W}_1,
\end{equation}

where $\mathcal{W}=W_p/nk_bT$ is the ratio of plasma wave energy density to thermal energy density of the plasma, and $\langle v_{ph}\rangle=\omega_p/\langle k_p\rangle$. A negative optical depth $\tau_{N1}$ indicates the possibility of an exponential dependence of the electromagnetic brightness temperature $T_b$ on the plasma wave energy density. 

The optical depth associated with collisional absorption of electromagnetic waves in the plasma wave generation region is: 
\begin{equation}
    \tau_{C}=\int_0^{R_s}\mu_{C}dl\approx\frac{6}{\sqrt 3}\frac{v_{th}\nu_{ei}}{c\langle v_{ph}\rangle}R_s
\end{equation}

As previously noted, maser amplification of electromagnetic radiation occurs when the induced conversion of plasma waves into electromagnetic waves dominates over collisional absorption, i.e., when the plasma wave energy density exceeds a certain threshold. This condition is derived from $|\tau_{N1}| \gg \tau_{C}$~\cite{Zaitsev2022}: 
\begin{equation}
  \mathcal{W}_1\gg\frac{108}\pi\frac{m_iv_T^2\nu_{ei}}{m_e\langle v_{ph}^2\rangle\omega_L} 
  \label{eq:condition_maser}
\end{equation}

The solution of the transfer equation (\ref{eq:transfer}) has the form:  
\begin{equation}
    T_b=\frac{a_i}{\mu_{Ni}+\mu_C}\left(1-\exp\left(-\tau_C-\tau_{Ni}\right)\right).
\end{equation}

Therefore, under condition (\ref{eq:condition_maser}), the expression for the observable radio flux $F$ resulting from Rayleigh scattering of plasma waves—generated by electrons accelerated upon reflection from the exoplanetary bow shock—takes the form: 
\begin{equation}
    F=3\frac{k_bTm_i}{c^2m_e}f_t^2\frac{R_s^2}{R_{SE}^2}\cdot\left(\exp\left(\frac{\pi}{18\sqrt{3}}\frac{m_e\omega_p\langle v_{ph}\rangle}{m_i cv_{th}}R_s\mathcal{W}_1\right)-1\right)\exp(-\tau_{ext}),
    \label{eq:flux_rl}
\end{equation}
where $\tau_{ext}$ accounts for collisional absorption of radio emission along the path from the source to the observer. The estimates below are made under the assumption $\tau_{ext} \ll 1$.

\subsection{Raman scattering}
\label{chapter5.2} 
In Raman scattering of light, electromagnetic waves are generated at the doubled plasma frequency $\omega_t=\omega_p^{(1)}+\omega_p^{(2)}\approx2\omega_p$ with a wave vector $\vec k_t\approx\vec k_p^{(1)}+\vec k_p^{(2)}$. The frequency $\omega_t$ and the wave number $\vec k_t$ of the electromagnetic wave are related by the dispersion relation $\omega_t^2=\omega_L^2+k_t^2c^2$. 

In the case of Raman scattering, an exponential dependence of the electromagnetic radiation intensity on the plasma wave energy density does not occur. As the electromagnetic field energy density increases, the reverse process becomes significant—namely, the decay of an electromagnetic wave at the doubled plasma frequency into two plasma waves with frequency $\omega_p$. This process is equivalent to effective scattering of electromagnetic waves~\cite{Zaitsev1983} and is described in the transfer equation~(\ref{eq:transfer}) by the nonlinear absorption coefficient $\mu_{N2}$. The collisional absorption coefficient $\mu_C$ is defined by the same expression~(\ref{eq:mu_C_1_coeff}) as in the case of Rayleigh scattering.

For an isotropic spectrum of plasma waves, the coefficients of spontaneous emission and nonlinear absorption in the transfer equation take the following forms~\cite{Zaitsev2023}: 
\begin{equation}
   a_2=\frac{\left(2\pi\right)^5}{15\sqrt 3}\frac{c^3}{\omega_L^2\langle v_{ph}\rangle}\frac{\mathcal{W}_2^2}{\xi^2}nT,~~    \mu_{N2}=\frac{\left(2\pi\right)^2}{5\sqrt 3}\frac{\omega_L}{\langle v_{ph}\rangle}\frac{\mathcal{W}_2}{\xi},
\end{equation}
where the parameter $\xi = c^3(\Delta k)^3/\omega_L^3$ characterizes the spectral volume of plasma waves. Assuming an isotropic plasma wave spectrum, the parameter $\xi$ can be estimated as:
\begin{equation}
    \xi\approx\frac{4\pi c^3}{\omega_L^3} \langle k_{p}\rangle^2\Delta k_p,
\end{equation}
where the quantities $\langle k_p \rangle$ and $\Delta k_p$ were estimated earlier~(\ref{eq:est_width_charakt}).

Focusing on the most relevant case where the plasma wave energy density is sufficiently high such that $\mu_{N2} \gg \mu_C$, and under the assumption of an optically thick source~\cite{Zaitsev2023}, one obtains the following estimate~(\ref{eq:flux_c}) for the radio flux observed on Earth, generated due to Raman scattering:
\begin{equation}
    F=\frac{2k_t^2k_b}{(2\pi)^2}\frac{R_s^2}{R_{se}^2}\frac{a_2}{\mu_{N2}}\exp(-\tau_{ext})=\frac{ k_t^2nk_bT}{3}\frac{R_s^2}{R_{se}^2}\frac{\mathcal{W}_2}{\langle k_{p}^2\rangle(k_{max}-k_{min})}\exp(-\tau_{ext}),
    \label{eq:flux_c}
\end{equation}

\subsection{Conversion into Radio Emission of Plasma Waves Generated by Electrons Accelerated at the Bow Shock of the Exoplanet HD~189733b}
\label{chapter5.3} 
\begin{figure}[h]
\includegraphics[width=8cm, height=6cm]{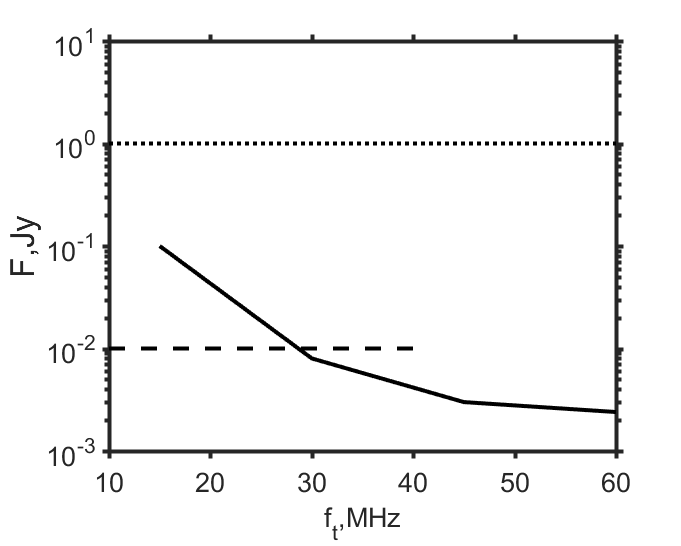}
\centering
\captionstyle{normal}
\caption{Approximate sensitivity of ground-based radio telescopes: LOFAR (solid line), UTR-2 (dashed line), and NDA (dotted line)~\cite{Griebmeier2011}.}
\label{ris:sensitivity}
\end{figure}

\begin{figure}[h]
\includegraphics[width=8cm, height=6cm]{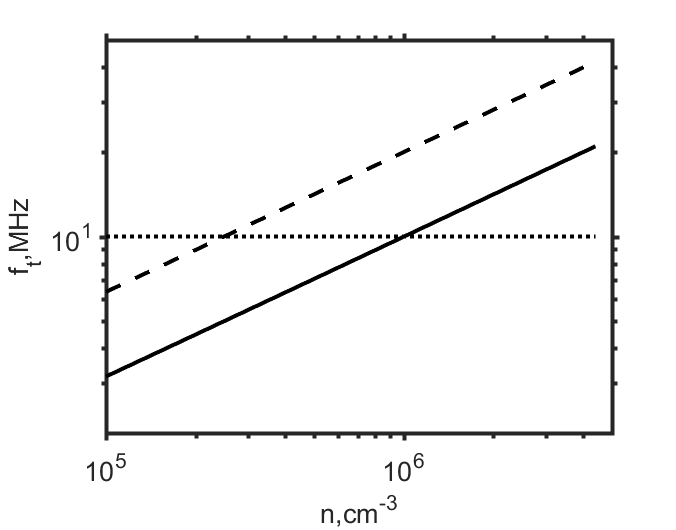}
\centering
\captionstyle{normal}
\caption{Approximate frequency of radio emission generated via Rayleigh (solid) and Raman (dashed) scattering of plasma waves as a function of stellar wind number density at temperature $T = 1.5 \cdot 10^6$ K. The ionospheric cutoff frequency is shown as a dotted line.}
\label{ris:frequency_border}
\end{figure}

The frequency of radio emission generated as a result of Rayleigh and Raman scattering is determined by the frequency of plasma waves, $\omega_p$, which slightly exceeds the Langmuir frequency $\omega_L$ and varies from 3 to 40 MHz (see Table~\ref{tab:results_en} and Fig.\ref{ris:frequency_border}). The most efficient radio astronomical instruments operating in this frequency range are UTR-2 (Ukrainian T-shaped Radio telescope-2)\footnote{Presumably damaged due to military actions in Ukraine. The authors do not possess reliable information about the current status of the telescope.}, LOFAR (LOw Frequency ARray), and NDA (Nançay Decameter Array). The sensitivity (i.e., the minimum radio flux density detectable with a 1-hour integration time and a 4 MHz bandwidth) of the UTR-2 radio telescope is 0.01 Jy in the 10–40 MHz band; for NDA, it is 1 Jy in the 10–120 MHz range (Fig.\ref{ris:sensitivity}). The sensitivity of LOFAR improves from 0.1 Jy to 0.005 Jy as the frequency increases from 15 to 40 MHz~\cite{Griebmeier2011}. Consequently, for further estimates, the detectable radio flux $F$ in Eqs.~(\ref{eq:flux_rl}) and (\ref{eq:flux_c}) is assumed to be 0.01 Jy throughout the entire frequency range.

The minimum energy of plasma waves $\mathcal{W}_1$, estimated from Eq.(\ref{eq:flux_rl}) to produce a radio flux of $F = 0.01$ Jy at Earth via Rayleigh scattering, is significantly lower than the energy of the accelerated electron beam for all four parameter sets (see Section~\ref{chapter3}) believed to be characteristic of stellar wind conditions near the bow shock of an exoplanet (cf. Table~\ref{tab:results} and Table~\ref{tab:results_en}). For parameter set N1, the ratio $\mathcal{W}_1/W$ is the lowest and approximately equal to 0.6. In the remaining cases, the required plasma wave energy $\mathcal{W}_1$ is 1–2 orders of magnitude lower than the estimated energy of the accelerated electrons. For parameter sets N1 and N3, the resulting radio emission frequencies are 3.2 MHz and 6.9 MHz, respectively, which fall below the ionospheric cutoff frequency of approximately 10 MHz~\cite{Griebmeier2011}. For the other two sets, the radio frequencies are around 20 MHz. Therefore, for stellar wind parameter sets N2 and N4, characterized by the highest wind density, the observation of radio emission resulting from Rayleigh scattering of plasma waves in the exoplanetary bow shock region is both energetically and spectrally feasible.

The ratio of the minimum plasma wave energy $\mathcal{W}_2$ (estimated using Eq.~(\ref{eq:flux_c})) required to produce a radio flux of $F = 0.01$ Jy via Raman scattering to the energy of the accelerated electrons, $\mathcal{W}_2/W$, is significantly below 1 only for parameter set N4, indicating that, for all other parameter sets, the detection of such emission from Earth is not feasible.

\begin{table}[h!]
\centering
\begin{tabular}{ |p{3cm}||p{2cm}|p{2cm}|p{2cm}|p{2cm}| }
 \hline
  Set & N1 & N2 & N3 & N4 \\
 \hline
  $f_p$, MHz& $3.2$& $21$  & $6.9$& $20$\\
   \hline
 $\mathcal{W}_1$& $0.16$ & $0.03$  & $0.06$ & $0.01$ \\
 \hline
 $\mathcal{W}_2$& $28$ & $0.14$  & $1.26$ & $0.16$ \\
 \hline
\end{tabular}
\caption{Properties of radio emission generated by accelerated electron beams for different parameter sets}
\label{tab:results_en}
\end{table}

\begin{figure}[h]
\includegraphics[width=8cm, height=6cm]{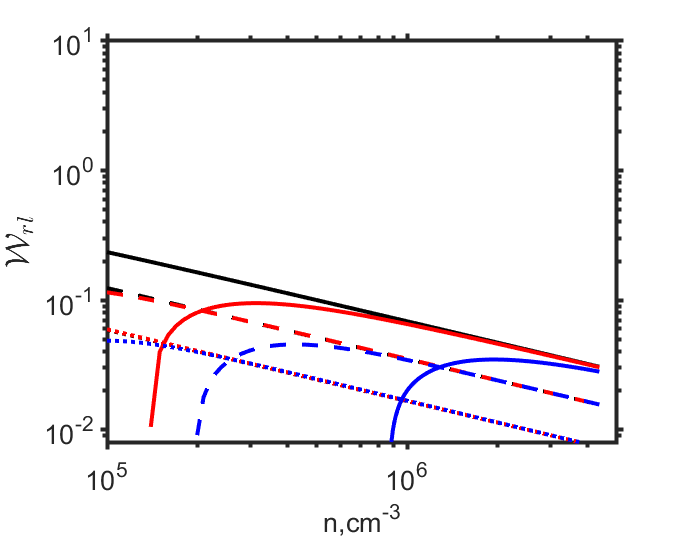}
\centering
\captionstyle{normal}
\caption{Dependence of the plasma wave energy density $\mathcal{W}_1$ required to generate a radio flux of $F = 0.01$ Jy on Earth via Rayleigh scattering, as a function of number density $n$ at temperature $T = 1.5 \cdot 10^6$ K, and magnetic field strengths $B = 0.01$ G (black), $B = 0.04$ G (red), $B = 0.1$ G (blue), for stellar wind velocities of $v = 250$ km/s (solid), $v = 500$ km/s (dashed), and $v = 1000$ km/s (dotted).}
\label{ris:W_rl}
\end{figure}

As previously discussed, the stellar wind along the orbit of HD~189733b is highly inhomogeneous. Therefore, it is important to consider how the potential for efficient generation of radio emission at the fundamental and second harmonic of the plasma frequency depends on stellar wind parameters. For stellar wind densities in the range of $10^5\text{cm}^{-3}$ to $5 \cdot 10^6\text{cm}^{-3}$ and a characteristic temperature of $T = 1.5 \cdot 10^6$ K, the plasma wave energy $\mathcal{W}_1$ required for efficient radio wave generation via Rayleigh scattering decreases with increasing magnetic field strength and stellar wind velocity. As the density increases, the dependence of plasma wave energy on the magnetic field disappears, and $\mathcal{W}_1$ becomes approximately inversely proportional to $\sqrt{n}$ (see Fig.\ref{ris:W_rl}). Thus, for an extremely fast stellar wind with $v_{sw} = 1000$ km/s, the condition for efficient radio wave generation, $W \gg \mathcal{W}1$, is already satisfied for densities $n \gtrsim 2 \cdot 10^5\text{cm}^{-3}$ and magnetic fields up to $0.1$ G. At $v_{sw} = 500$ km/s, the same holds for magnetic fields up to 0.04 G, and for $v_{sw} = 250$ km/s, the condition is only met for $n \gtrsim 3 \cdot 10^6\text{cm}^{-3}$ (cf. Figs.\ref{ris:W_zav_B2_V2} and \ref{ris:W_rl}). It is also worth noting that, due to the maser effect in Rayleigh scattering, the required plasma wave energy density $W \gg \mathcal{W}_1$ depends only logarithmically on the minimum detectable radio flux $F$(\ref{eq:flux_rl}), so a reduction in sensitivity to 1 Jy (as with the NDA telescope) does not significantly alter the conditions for efficient generation.

\begin{figure}[h]
\includegraphics[width=8cm, height=6cm]{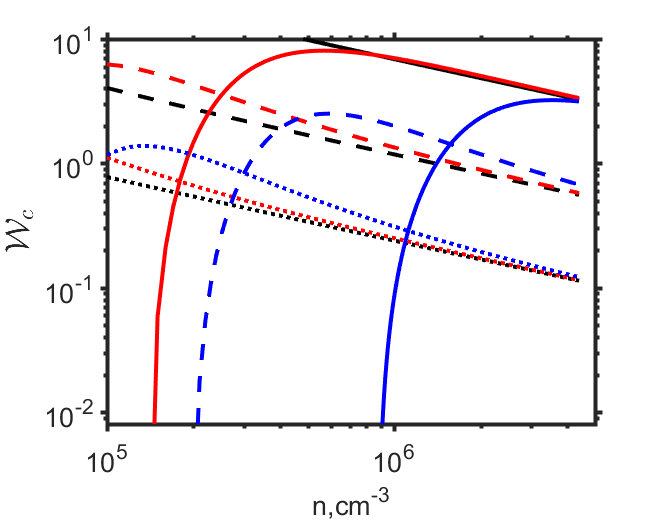}
\centering
\captionstyle{normal}
\caption{Dependence of the plasma wave energy density $\mathcal{W}_2$ required to generate a radio flux of $F = 0.01$ Jy on Earth via Raman scattering, as a function of number density $n$ at temperature $T = 1.5 \cdot 10^6$ K, and magnetic field strengths $B = 0.01$ G (black), $B = 0.04$ G (red), $B = 0.1$ G (blue), for stellar wind velocities of $v = 250$ km/s (solid), $v = 500$ km/s (dashed), and $v = 1000$ km/s (dotted).}
\label{ris:W_c}
\end{figure}

In the considered stellar wind parameter space, the plasma wave energy $\mathcal{W}_2$ required for efficient radio wave generation via Raman scattering is significantly higher than that required for Rayleigh scattering. Thus, the condition $W \gg \mathcal{W}_2$ is satisfied only at extremely high stellar wind velocities $v_{sw} \gtrsim 1000$ km/s and densities $n \gtrsim 10^6\text{cm}^{-3}$ (cf. Figs.~\ref{ris:W_zav_B2_V2} and \ref{ris:W_c}).

When discussing the possibility of detecting radio emission at Earth, one must account for the fact that the ionosphere absorbs radio waves with frequencies below 10 MHz. Figure~\ref{ris:frequency_border} shows that radiation at the fundamental plasma frequency can only penetrate the ionosphere for stellar wind densities $n \gtrsim 10^6\text{cm}^{-3}$, while at the second harmonic, it becomes detectable for $n \gtrsim 2.5 \cdot 10^5\text{cm}^{-3}$.

Thus, due to spectral and energetic limitations, detection of radio emission from the bow shock of exoplanet HD~189733b at Earth's surface is only feasible if the stellar wind density exceeds $n \gtrsim 10^6;\text{cm}^{-3}$. The emission frequency is approximately equal to the fundamental plasma frequency, which is primarily determined by the density and varies from 10 to 20 MHz under typical stellar wind conditions. For extremely fast stellar winds with $v_{sw} \gtrsim 1000$ km/s, efficient generation of radio waves at the second harmonic may also be achievable with sufficient strength for ground-based detection.

\section{Conclusion}

This study investigates the capabilities of the shock drift acceleration mechanism in the context of a quasi-perpendicular bow shock associated with the exoplanet HD~189733b. In particular, we provide estimates of the density, velocity, and energy density of the beam of accelerated electrons, which, under characteristic stellar wind parameters in this system, prove to be sufficiently energetic to excite plasma waves.

On the other hand, the study also examines the efficiency of the plasma mechanism for radio wave generation in the stellar wind near the exoplanetary bow shock of HD~189733b. In this mechanism, the primary energy source is a population of energetic electrons that excite plasma waves, which are subsequently converted into electromagnetic radiation through scattering—either at the fundamental plasma frequency via Rayleigh scattering or at the second harmonic frequency via Raman scattering. Therefore, comparing the energy of the accelerated electrons with the minimum plasma wave energy required for the resulting electromagnetic radiation to be detectable by modern ground-based radio telescopes allows us to estimate the region of stellar wind parameters where radio emission from the exoplanetary bow shock is energetically observable.

The efficiency of radio wave generation increases with rising stellar wind density and velocity, and decreases with increasing magnetic field strength. The analysis shows that, for typical values of the magnetic field, temperature, and stellar wind velocity in the HD~189733 system, radio wave detection at the fundamental plasma frequency requires the local stellar wind density near the bow shock to exceed $n\gtrsim10^6,\textrm{cm}^{-3}$. Moreover, if the stellar wind velocity exceeds $v_{sw}\gtrsim1000$~km/s, radio wave detection at the second harmonic plasma frequency also becomes energetically feasible. The most promising frequency range for radio detection lies near the ionospheric cutoff.

This study does not account for numerous factors that may either enhance or suppress the radio emission from the exoplanetary bow shock. Such factors include: refined estimates of the size, location, shape, and type of the exoplanetary bow shock; possible radio wave refraction within the HD~189733 system; gyroabsorption~\cite{Stepanov1999}; collisional absorption during wave propagation from the source to the observer; and differences between ion and electron temperatures in the stellar wind near the shock front. Furthermore, the complete velocity distribution function of electrons obtained in this work (equation~\ref{eq:full_fr}) is unstable not only with respect to Langmuir modes but also to filamentation-type Weibel perturbations~\cite{Weibel1959,Bret2004,Kuznetsov2023}. The development of such instabilities may significantly influence plasma wave generation and, consequently, the resulting radio emission. These effects may differentially impact the intensity of radio waves at the fundamental and second harmonic plasma frequencies, suggesting that both Rayleigh and Raman scattering mechanisms could generate detectable radio emission. These effects, therefore, warrant detailed investigation in future studies.

An interesting avenue for future studies could be the application of other known acceleration mechanisms, such as the surfatron mechanism~\cite{Kichigin1995}, to plasma in the region of an exoplanetary bow shock

A potential detection of radio emission could provide direct insight into the properties of the exoplanetary bow shock and the parameters of the stellar wind in its vicinity. This would significantly refine existing gas-dynamical and magnetohydrodynamical models of the HD~189733 system and improve our understanding of the interaction between hot Jupiters and their host stars.

This research was supported by the Theoretical Physics and Mathematics Advancement Foundation “BASIS” (project no. 24-1-5-94-1 and (project no. 24-1-1-97-1)
\printbibliography
\end{document}